\documentclass[amsmath, twocolumn, showpacs, aps, prl]{revtex4-1}
\usepackage{bm}
\usepackage{graphics}
\usepackage{graphicx}

\begin{document}

\title{Correlation effects in disordered conductors with spin accumulation}
\author{A. A. Zyuzin$^{1,2}$ and A. Yu. Zyuzin$^1$}

\affiliation{ $^1$ A.F. Ioffe Physico-Technical Institute of
Russian Academy of Sciences, 194021 St. Petersburg, Russia\\ $^2$
 Department of Physics, University of Basel,
Klingelbergstrasse 82, CH-4056 Basel, Switzerland }


\begin{abstract}
We consider the effect of electron-electron interaction on the
density of states of disordered paramagnetic conductor
in the presence of spin accumulation and magnetic field.
We show that interaction correction to electron density of states
of the paramagnet may exhibit singularities at energies
corresponding to the difference between chemical potentials of
electrons with opposite spins.
We also discuss correlation effects on conductivity in metallic as well as in hopping regimes and
show that spin accumulation leads to the negative magnetoconductivity.
\end{abstract}
\maketitle

\section{Introduction}

The realization of large spin accumulation, that can be described by introducing the quasi-chemical 
potentials $\mu\pm\delta\mu/2$ for
spin-up and spin-down electrons in the system possessing long spin relaxation time, is important for many proposals in spintronics. 
In particular, the non-equilibrium spin polarization might be created by illuminating the sample with
circularly polarized light \cite{bib:Optical}, or can be achieved in conductors placed in contact to the ferromagnet
via the spin injection mechanism \cite{bib:Aronov, bib:JS}, see for a review
\cite{bib:Dyakonov1, bib:Review}. The spin injection
and detection were investigated in many works and for
different types of materials such as superconductors
\cite{bib:supercond, bib:supercond1, bib:supercond2}, organic
polymers \cite{bib:polymers-review}, graphene
\cite{bib:inj-graphene}. Recently there has been a progress in
achieving of large spin accumulation by means of electrical spin injection
in semiconductors such as $\textrm{GaAs}$, $\textrm{Si}$, $\textrm{Ge}$ \cite{bib:Review-Si2,
bib:Review-Si3, bib:Review-Si}.

Motivated by experiments on spin accumulation, we focus on the
non-equilibrium correlation effects in conductivity and density of states (DOS) in the presence of spin accumulation.

Quantum corrections to the transport and
thermodynamical properties of disordered metallic conductors have been a
subject of both theoretical and experimental studies. Electron-electron
interaction in disordered conductors results in the singularities
of electron DOS at the Fermi level and positive magnetoresistivity, for a review see \cite{bib:AA, bib:Lee}.

For example,
the energy dependence of the interaction correction to the
electron DOS at zero temperature has a logarithmic singularity in two
dimensions $\delta\nu(\epsilon)\propto \ln|\epsilon\tau|$, where
$\tau$ is the electron mean free time. In addition, external
magnetic field $B$ due to Zeeman splitting leads to singularities shifted from the Fermi energy by the amount $\pm\Omega_z$ as
$\delta\nu(\epsilon)\propto \ln|(\epsilon^2-\Omega^{2}_z)\tau^2|$, where
$\Omega_z =\lvert g\mu_B  B\rvert \mathrm{sign} (B)$ and $\mu_B$ is the
Bohr magneton. We define $\Omega_z$ in such a way that it can be of either sign, depending on the direction 
of magnetic field. The interaction correction to the conductivity decreases in
magnetic field in the limit of weak spin relaxation as \cite{bib:Lee} $ \delta\sigma\sim-\lambda\nu\ln \lvert\Omega_z\tau\rvert$, where $\lambda$ is the electron-electron interaction constant and $\nu$ is DOS at the Fermi level per one spin.
More detailed discussion can be found in \cite{bib:AA, bib:Lee}.

Correlation effects related to the spin degree of freedom can also play a significant role in magnetoresistance in the
hopping conductivity regime. If the Coulomb repulsion between electrons permits the onsite double occupancy, then the probability of 
certain transitions under the Zeeman splitting decreases leading to positive magnetoresistivity \cite{bib:Kamim, bib:Kamim1, bib:Matv}.

In the following sections we consider correlation effects in DOS and conductivity in case of finite spin accumulation $\delta\mu$. 
We essentially rely on the calculations done for the equilibrium case, therefore we discuss only the features that 
distinguish the non-equilibrium case.

\section{Density of states in metallic region}
We consider disordered metallic conductor in the presence of spin accumulation and magnetic field. The spin accumulation can be
obtained either by spin injection or optical orientation methods. We assume that the spin relaxation time in the system 
is much longer than the energy relaxation time such that we
can treat the system in the energy equilibrium regime while having the non-equilibrium spin polarization.
Within this assumption the interaction correction to the one particle DOS in the non-equilibrium state generated by spin accumulation
can be treated in the similar way done for the equilibrium case \cite{bib:AA}.  We need to take into
account the non-equilibrium exchange splitting in the definition of retarded and advanced Green functions and to modify 
the Fermi distribution function of electrons by introducing the chemical potential shifts of electrons with spin-up and spin-down.

\begin{figure}[t]  \centering
\includegraphics[width=8cm] {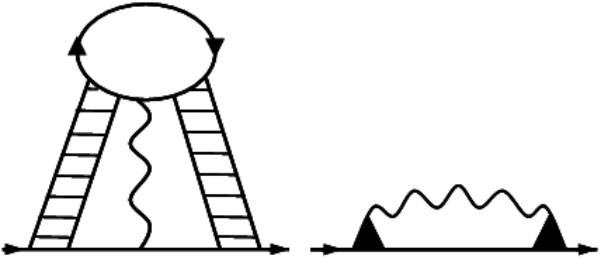}
\caption{Diagram for the calculation of the density of states.
}\label{fig:1}
\end{figure}

Interaction correction to the one particle DOS is determined by the advanced component of Green function
\begin{equation}
 \delta\nu_{\alpha}(\epsilon)=\frac{1}{\pi}\mathrm{Im}\int\frac{d\mathbf{p}}{(2\pi)^3} \delta G_{A}(\mathbf{p},\epsilon,\alpha)
\end{equation}
where $\alpha=\pm $ defines the spin direction and we accept the units $\hbar\equiv 1$.
Let us consider the exchange part of the interaction correction shown in Fig. (\ref{fig:1}), 
where diffusion contribution to the self-energy part is given as
\begin{eqnarray}\label{sigma}\nonumber
\Sigma _{A}\left(\mathbf{p},\epsilon,\alpha\right) =G_{R}\left(\mathbf{p},0,\alpha\right)\frac{1}{2V}\sum_{\mathbf{q}}
\int\frac{d\omega}{2\pi}\times\\ \times \left[ F(\epsilon,\alpha)-F(\omega,\alpha)\right]
 D_{0}^{2} \left(\omega-\epsilon,\mathbf{q}\right) V_{A}\left(\omega-\epsilon, \mathbf{q}\right)
\end{eqnarray}
here $D_{0}\left(\omega,\mathbf{q}\right)=(D\mathbf{q}^2-i\omega)^{-1}$ is the diffusion propagator, $D$ is the diffusion coefficient, $V$ is the volume of the system
and screened Coulomb potential is given by the following expression
\begin{equation}
V_{A}(\omega,\mathbf{q}) =
\frac{V_0(\mathbf{q})}{1+2\nu V_0(\mathbf{q})
\Pi_0(\omega,\mathbf{q})}
\end{equation}
where $V_0(\mathbf{q})$ is the Coulomb potential and $\Pi_0(\omega,\mathbf{q})=D\mathbf{q}^2/(D\mathbf{q}^2-i\omega)$.
Function $F(\omega,\alpha)$ is determined by the Keldysh component of the non-equilibrium Green function, which after 
averaging over random potential becomes
\begin{equation}
G_{K}\left(\mathbf{p},\omega,\alpha\right) =\frac{1}{i\tau }
G_{R}\left(\mathbf{p},\omega,\alpha\right) F\left(\omega,\alpha\right) G_{A}\left(\mathbf{p},\omega,\alpha
\right)
\end{equation}

where $G_{R/A}\left(\mathbf{p},\omega,\alpha\right)$ are the retarded/advanced Green functions of electrons
with spin direction $\alpha=\pm$.
Function $F(\omega,\alpha)$ in the uniform state described by the chemical potential shifts $\delta\mu$ can be written as
\begin{equation}\label{udistr}
F\left(\omega,\alpha\right)=1-2n_{\alpha}(\omega)=\tanh\left(\frac{\omega+\alpha\delta\mu/2}{2T}\right)
\end{equation}
We see that the integral over $\omega$ of the first term in expression
(\ref{sigma}) proportional to $F(\epsilon,\alpha)$ is equal to zero since all diffusion poles are in the region $\mathrm{Im}(\omega)<0$.
Substituting (\ref{udistr}) into (\ref{sigma}) we note that expression (\ref{sigma}) is equal to the equilibrium
one under the substitution of energy $\epsilon-\alpha\delta\mu/2$ by energy $\epsilon$ in
the expression for correction to DOS for spin direction $\alpha$. Therefore, the total correction to DOS for spin direction $\alpha$ has the form
\begin{widetext}
\begin{equation}
\delta\nu_{\alpha}(\epsilon)=-\frac{\nu}{V}\mathrm{Im}\sum_{\mathbf{q}}\int
\frac{d\omega}{2\pi}
\left[\frac{U_{R,\alpha}(\omega,\mathbf{q})\tanh\left(\frac{\epsilon-\omega+\alpha\delta\mu/2}{2T}\right)}{[D\mathbf{q}^2-i\omega-
i \alpha(\Omega_z + \Omega_p)]^2}+
\frac{1}{2}\frac{(U_{R,0}(\omega,\mathbf{q})-V_{R}(\omega,\mathbf{q}))\tanh\left(\frac{\epsilon-\omega-\alpha\delta\mu/2}{2T}\right)}{[D\mathbf{q}^2-
i\omega]^2}\right]
\end{equation}
\end{widetext}
where we define
\begin{equation}
U_{R,M}(\omega,\mathbf{q}) = \frac{\lambda}{1-\lambda\nu
\widehat{\Pi}_M(\omega,\mathbf{q})}
\end{equation}
and $M=(0,\alpha$), while the polarization operator is
\begin{equation}
\widehat{\Pi}_M(\omega,\mathbf{q}) =
\frac{D\mathbf{q}^2-iM\Omega_p/\lambda\nu}{D\mathbf{q}^2-i\omega-iM(\Omega_z+\Omega_p)}
\end{equation}
Here we include Zeeman splitting $\Omega_z$ and exchange energy that is self-consistently defined as
\begin{equation}
\Omega_p=\frac{\lambda}{2}\int \frac{d\epsilon}{2\pi}\nu(\epsilon)[F(\epsilon,+)-F(\epsilon,-)]=\frac{\lambda\nu}{1-\lambda\nu}[\Omega_z+\delta\mu]
\end{equation}
We assume uniform spin accumulation and magnetic field applied
parallel to the non-equilibrium magnetization.
Magnetic field polarizes electron spin in the system
and as a result the spin polarization $\mathcal{S}$ in the paramagnet is a sum of
two contributions coming from spin accumulation and applied
magnetic field.
\begin{equation}\label{magnetiz}
\mathcal{S}=\frac{\nu}{1-\lambda\nu}[\Omega_z+\delta\mu]
\end{equation}
Note, that we always assume $\delta\mu$ to be positive. At the same 
time $\Omega_z$ can be either positive or negative depending on direction of the magnetic field.

Let us now consider spin accumulation in the paramagnet in the two-dimensional limit. We take into account that 
$V^{-1}\sum_{\mathbf{q}}\rightarrow\int\frac{d^2\mathbf{q}}{(2\pi)^2}$, 
Coulomb interaction is $V_0(\mathbf{q})=e^2/q$ and we also assume small parameter $\lambda\nu<1$.
We find that the energy dependence of the interaction correction to
the density of states in two
dimensional conductor at $|\epsilon|,\delta\mu>T$ has the form
\begin{widetext}
\begin{equation}\label{updown}
\delta\nu_{\uparrow,\downarrow}(\epsilon)=-\frac{1}{4\pi^2D}\biggl(
\frac{1}{4}\ln\left|\frac{\epsilon\mp\delta\mu/2}
{D^2\kappa^4\tau}\right|\ln(\left|\epsilon\mp\delta\mu/2\right|\tau)
+
\frac{\lambda\nu}{2}\ln(\left|\epsilon\mp\delta\mu/2\right|\tau)+\lambda\nu\ln(\left|\epsilon\pm(\Omega_z+\Omega_{p}+\delta\mu/2)\right|\tau) \biggr)
\end{equation}
\end{widetext}
where  $\kappa = 2\pi e^2
\nu$ is the inverse static screening length.

Comparing this expression with that in the equilibrium case \cite{bib:AA},
we find that when spin polarization $\mathcal{S}$ is zero, non-equilibrium spin accumulation splits the
singularities of the density of states at $\epsilon=0$  \cite{bib:AA} to
$\epsilon=\pm \delta\mu/2$ for electron with spin up and down, correspondingly. If $\mathcal{S}\neq 0$ then the density of states exhibits singularities 
located at
$\epsilon=\pm |\delta\mu/2 + \Omega_z|$ for small $\lambda\nu$ in addition to the singularities at $\pm \delta\mu/2$. This case is similar 
to the equilibrium one in the presence of magnetic field.

\section{conductivity}
\subsection{Metallic region}
When considering the interaction correction to conductivity in the presence of chemical potential shifts we can use the 
results obtained for equilibrium \cite{bib:AA, bib:Lee} similarly as it was done for DOS. The combined effect of 
external magnetic field and spin accumulation on conductivity is determined by total spin polarization. 
Expression for conductivity contains function $\frac{d}{d\omega}[\omega\coth(\omega/2T)]$ that appears as a result of 
integration of the equilibrium Fermi distribution functions.
Instead of these equilibrium functions we obtain for $\delta\mu$-dependent part of conductivity 
\begin{eqnarray}\nonumber
-\frac{1}{2}\frac{d}{d\omega}\int^{\infty}_{-\infty} d\epsilon \left[F\left(\epsilon,\alpha\right)F\left(\epsilon+\omega,-\alpha\right)-1\right]=\\=\frac{d}{d\omega}(\omega-\alpha\delta\mu)\coth\left(\frac{\omega-\alpha\delta\mu}{2T}\right)
\end{eqnarray}
As a result, we obtain the expression for the spin accumulation dependent interaction correction to the conductivity
\begin{widetext}
\begin{equation}\label{s1}
\delta\sigma =-i\frac{2\sigma _{d}}{\pi d}\frac{1}{V}\sum_{\mathbf{q},\alpha=\pm}D\mathbf{q}^{2}\int d\omega\frac{U_{R,\alpha}(\omega,\mathbf{q})\partial_{\omega}\left[(\omega-\alpha\delta\mu) \coth \frac{(\omega-\alpha\delta\mu)}{2T}\right]}{\left( D\mathbf{q}^{2}-i\omega-i\alpha(\Omega_z + \Omega_p) \right) ^{3}}
\end{equation}
\end{widetext}
Expression (\ref{s1}) coincides with that in the equilibrium case \cite{bib:Lee} with the corresponding spin polarization $\mathcal{S}$, 
similarly as it was done for DOS .
The interaction 
correction to the conductivity of two-dimensional metallic system in the limit of large spin accumulation $\delta\mu>T$ becomes
\begin{equation}\label{weakCond}
 \sigma(\delta\mu)-\sigma(0)=-\frac{\lambda\nu e^2}{4\pi^2\hbar}\ln|\delta\mu\tau|
\end{equation}
while in the small spin accumulation $\delta\mu<T$ regime the interaction correction takes the form
\begin{equation}\label{strongCond}
\sigma(\delta\mu)-\sigma(0)= -0.08\frac{\lambda\nu e^2}{4\pi^2\hbar}\left(\frac{\delta\mu}{T}\right)^2
\end{equation}
If the external magnetic field is applied to the system then $\delta\mu$ in expressions 
(\ref{weakCond}) and (\ref{strongCond}) must be substituted with 
$\lvert\delta\mu+\Omega_{z}\rvert$. 
However, when considering the frequency dependent conductivity the equivalence
between $\delta\mu$ and Zeeman energy will be lost.

\subsection{Hopping region}
Let us consider hopping conductivity in the presence of spin accumulation.
Spin dependent contribution to the hopping conductivity 
arises if electron hopping involves states that permit double onsite
occupancy in the presence of Coulomb repulsion \cite{bib:Kamim, bib:Kamim1}.
According to \cite{bib:Kamim, bib:Kamim1} hopping rates are determined by correlation functions of occupation numbers of the sites 
involved in hopping.
The one site density matrix is given as
\begin{equation}\label{DMATR}
 \varrho \sim\exp\left(-\frac{\epsilon_{i}(n_{+}+n_{-})}{T}+\frac{Un_{+}n_{-}}{T}+\frac{\delta\mu(n_{+}-n_{-})}{2T}\right)
\end{equation}
Here $\epsilon_{i}$ is the one electron energy of 
localized state $i$, counted from the Fermi level and $U>0$ is the intra-site electron repulsion potential.

Expression (\ref{DMATR}) coincides with equilibrium one for electrons in magnetic field.
After mapping the approach developed in \cite{bib:Kamim, bib:Kamim1} to the case of spin accumulation the
hopping rate from site $i$ to site $k$ can be written as sum of contributions proportional to
\begin{enumerate}
\item $\exp(-[\epsilon_{i}\pm\delta\mu/2]/T)/Z_{i}Z_{k}$ if site $i$ has one electron and site $k$ is empty,
\item  $\exp(-[\epsilon_{i}+\epsilon_{k}]/T)/Z_{i}Z_{k}$ if both sites are single occupied,
\item $\exp(-[2\epsilon_{i}+U]/T)/Z_{i}Z_{k}$ if site $i$ has two electrons and site $k$ is empty,
\item  $\exp(-[2\epsilon_{i}+\epsilon_{k}+U\mp\delta\mu/2]/T)/Z_{i}Z_{k}$ if site $i$ has two electrons and site $k$ has one electron.
\end{enumerate}
where partition function in the case of chemical potential shifts is given as
\begin{equation}
Z_{i}=1+2\cosh\left(\frac{\delta\mu}{2T}\right)e^{-\epsilon_{i}/T}+e^{-(2\epsilon_{i}+U)/T}
\end{equation}
for $U>0$ assumed to be equal for all sites.

Scaling expression for the variable range hopping conductivity in this problem was 
obtained in \cite{bib:Matv}. For small values of spin accumulation defined by the inequality $\frac{\delta\mu}{T}(\frac{T}{T_{0}})^{1/(d+1)}<1$, where $T_{0}$ 
is the characteristic Mott temperature and $d$ is the sample dimension, we obtain the expression 
for the spin accumulation $\delta\mu$ dependent part of conductivity
\begin{equation}\label{hopp}
 \ln\left|\frac{\sigma(\delta\mu)}{\sigma(0)}\right|=-\frac{g_{S}g_{D}}{g_{S}+g_{D}}\frac{\delta\mu}{T}
\end{equation}
where $g_{S}$ and $g_{D}$ are the densities of states with energies $\epsilon_{\alpha}$ and $\epsilon_{\alpha}+U$ 
near the Fermi level correspondingly.
Note that in this regime $\ln|\sigma(\delta\mu)/\sigma(0)|$ does not depend on dimension $d$.
Again, expression (\ref{hopp}) has to be modified in the presence of magnetic field. 
Spin accumulation term $\delta\mu$ must be substituted with $\lvert\delta\mu+\Omega_{z}\rvert$.

\section{Conclusions}
The singularities of DOS can be measured with a tunnelling probe electrode, 
where the singularities can reveal in the zero bias dips of the tunnelling conductivity \cite{bib:AA}.  We note that in spin 
accumulation regime both sides of the tunnelling contact will acquire chemical potential shifts. In order to resolve 
the splitting of singularities in DOS, the chemical potential splitting in the probe electrode has to be much smaller 
than that in the studied system. This can by achieved if one takes the probe electrode with much stronger spin relaxation 
compared to the spin relaxation in the studied system.

Let us now discuss the experimental observability of considered correlation
effects. An important
issue is the value of chemical potential shift $\delta\mu$
compared to the value of the inverse spin relaxation time $1/\tau_s$ in the system. The
applicability of expressions (\ref{updown}), (\ref{weakCond}) and (\ref{strongCond}) requires
$\sqrt{\delta\mu\tau_s}$ to be large. We take the spin relaxation
time to be $\tau_s\sim 100$ ps and the spin accumulation
$\delta\mu \sim 1$ meV. The $1$ meV spin accumulation requires
correspondingly low $<10$ K temperatures. We obtain
$\sqrt{\delta\mu\tau_s} \sim 10$. Also assuming the diffusion
coefficient to be $10$ cm$^2$ s$^{-1}$, we estimate $\ell_s
=\sqrt{D\tau_s} \sim 300$ nm and $\sqrt{D/|\delta\mu|} \sim 30$
nm. The above conditions on the magnitude of the spin relaxation
length and spin accumulation are realizable experimentally in the
ferromagnet - doped semiconductors contacts \cite{bib:Review-Si,
bib:Estim1, bib:Estim2, bib:Estim3}. However, the nature of this
large value of spin accumulation is still under debate
\cite{bib:Tran, bib:Jansen}.

We are grateful for the financial support of RFFI under Grant No.
12-02-00300-A.

\end{document}